\documentclass[twocolumn]{IEEEtran}
\usepackage{times}

\usepackage{amsmath}  
\usepackage{amssymb}  
\usepackage{mathrsfs} 

\usepackage{theorem}  
\usepackage{cite}     
\usepackage{comment}  

\usepackage{upref}
\usepackage{amsfonts}

\usepackage{verbatim}

\usepackage[dvipsnames,usenames]{color}
\usepackage{enumerate}

\usepackage{graphicx}
\usepackage{subfigure}

\usepackage{latexsym}
\usepackage[normalem]{ulem}

\newcommand{\be}[1]{\begin{equation}\label{#1}}
\newcommand{\ee}{\end{equation}}

\newcommand{\bc}{\begin{center}}
\newcommand{\ec}{\end{center}}

\newcommand{\cC}{{\cal C}}
\newcommand{\cD}{{\cal D}}

\newcommand{\cI}{{\cal I}}

\newcommand{\cU}{{\cal U}}

\newcommand{\cX}{{\cal X}}

\newcommand{\bfa}{{\boldsymbol a}}

\newcommand{\bfu}{{\boldsymbol u}}
\newcommand{\bfv}{{\boldsymbol v}}
\newcommand{\bfw}{{\boldsymbol w}}
\newcommand{\bfx}{{\boldsymbol x}}
\newcommand{\bfy}{{\boldsymbol y}}
\newcommand{\bfz}{{\boldsymbol z}}

\newcommand{\bfU}{{\mathbf U}}
\newcommand{\bfV}{{\mathbf V}}
\newcommand{\bfW}{{\mathbf W}}
\newcommand{\bfX}{{\mathbf X}}
\newcommand{\bfY}{{\mathbf Y}}
\newcommand{\bfZ}{{\mathbf Z}}

\renewcommand{\leq}{\leqslant}

\renewcommand{\geq}{\geqslant}

\newcommand{\Cref}[1]{Co\-rol\-la\-ry\,\ref{#1}}

\newcommand{\nchoosek}[2]{\left(\begin{array}{c}#1\\#2\end{array}\right)}

\theoremstyle{plain} \theorembodyfont{\normalfont\slshape}

\newtheorem{thm}{Theorem$\!$}
\newenvironment{theorem}{\begin{thm}\hspace*{-1ex}{\bf.}}{\end{thm}}

\newtheorem{prop}[thm]{Proposition$\!$}

\newtheorem{lem}[thm]{Lemma$\!$}
\newenvironment{lemma}{\begin{lem}\hspace*{-1ex}{\bf.}}{\end{lem}}

\newtheorem{cor}[thm]{Corollary$\!$}
\newenvironment{corollary}{\begin{cor}\hspace*{-1ex}{\bf.}}{\end{cor}}

\newtheorem{prob}[thm]{Problem$\!$}

\newtheorem{defi}[thm]{Definition$\!$}

\newtheorem{claim}{Claim}

\theorembodyfont{\normalfont}

\newtheorem{exam}{Example$\!$}
\newenvironment{example}{\begin{exam}\hspace*{-1ex}{\bf .}}{\end{exam}}

\newtheorem{remrk}{Remark$\!$}

\definecolor{Codecolor}{named}{White}  

\newcommand{\Copen}{\mbox{\{\kern-5.50pt\{}}
\newcommand{\Cclose}{\mbox{\}\kern-5.50pt\}}}
\newcommand{\Cslash}{\mbox{$\backslash\kern-6.02pt\backslash$}}

\begin{document}
\title{The Hybrid $k$-Deck Problem: Reconstructing Sequences from Short and Long Traces}
\author{
  \IEEEauthorblockN{
    Ryan~Gabrys\IEEEauthorrefmark{1}\IEEEauthorrefmark{2}~and
    Olgica~Milenkovic\IEEEauthorrefmark{1}}
  {\normalsize
    \begin{tabular}{ccc}
      \IEEEauthorrefmark{1}ECE Department, University of Illinois, Urbana-Champaign~~ &
      \IEEEauthorrefmark{2}Spawar Systems Center, Pacific~~ \\
    \end{tabular}}\vspace{-3ex}
    }

\maketitle

\begin{abstract} We introduce a new variant of the $k$-deck problem, which in its traditional formulation asks for determining the smallest $k$ that allows one to reconstruct any binary sequence of length $n$ from the multiset of its $k$-length subsequences. In our version of the problem, termed the hybrid $k$-deck problem, one is given a certain number of special subsequences of the sequence of length $n-t$, $t>0$, and the question of interest is to determine the smallest value of $k$ such that the $k$-deck, along with the subsequences, allows for reconstructing the original sequence in an error-free manner. We first consider the case that one is given a single subsequence of the sequence of length $n-t$, obtained by deleting zeros only, and seek the value of $k$ that allows for hybrid reconstruction. We prove that in this case, $k \in [\log t+2, 
\min\{{t+1,O( \sqrt{n \cdot (1+\log t)})\}}]$. We then proceed to extend the single-subsequence setup to the case where one is given $M$ subsequences of length $n-t$ obtained by deleting zeroes only. In this case, we first aggregate the asymmetric traces and then invoke the single-trace results. The analysis and problem at hand are motivated by nanopore sequencing problems for DNA-based data storage. 
\end{abstract}

\section{Introduction}

The $k$-deck of a sequence $\bfx$ of length $n$ is the multiset of all its subsequences of length $k$. A sequence that is uniquely defined by its $k$-deck is termed $k$-deck reconstructable. The $k$-deck problem is to determine $f(n)$, the smallest value of $k$ such that any sequence $\bfx$ of length $n$ is reconstructable from its $k$-deck. The problem was first described in~\cite{kalashnik73}, where it was also shown that $f(n) \leq \lfloor n/2 \rfloor$. The first lower bounds were 
established in~\cite{zenkin1984}, and improved bounds were described in~\cite{KR73} and \cite{Scott}. The $k$-deck problem is also closely related to a number of other reconstruction problems that have received significant attention, such as trace reconstruction~\cite{batu2004}, reconstruction of graphs from subgraphs~\cite{bondy77}, and set reconstruction based on multiset information~\cite{acharya2015}.

The $k$-deck problem may be viewed as an abstracted version of a DNA nanopore sequencing problem~\cite{oxford}. In this context, a string is passed through the nanopore multiple times, and at each pass a \emph{trace sequence} is produced. Sequencing traces arise due to insertions, deletions and substitution edits in the original sequence and are usually of variable length. For simplicity, we consider traces obtained via deletions only, all of which have the same length. One issue in nanopore sequencing that was observed in the experimental study of the authors~\cite{yazdi2016} is that the biological ``nanopore channels'' tend to degrade in time: The sequences produced in the first hour of sequencing usually contain fewer errors (i.e., fewer deletions) and are hence of longer length than the sequences produced later in the process. Furthermore, early deletion errors appear to be context dependent, in so far that so called purine symbols (bases) show larger error rates than pyrimidine symbols\footnote{The DNA bases $A$ and $G$ are called purines, while $T$ and $C$ are called pyramidines.}. We abstract this observation by assuming that the ``good'' sequencing channels are asymmetric, in so far that they delete only purines. In this case, it suffices to focus on analyzing binary sequences only, as ``0'' may be used to designate purines, and ``1'' may be used to designate pyrimidines. 

The above discussion motivates the introduction of a ``hybrid'' sequence reconstruction problem, in which one is given a small set of long (length $n-t$, $t>0$), asymmetric subsequences of a sequence $\bfx$, and asked to determine the shortest length of a large set of shorter (length $k$) subsequences that allows for unique reconstruction of $\bfx$. We refer to this problem as the \emph{hybrid $k$-deck problem}. Our results on the hybrid $k$-deck problem include lower and upper bounds on the smallest $k$ that allows for exact sequence reconstruction, for the case that only one asymmetric sequence of length $n-t$ is given, or for the case that $M$ such sequences are available. A related, simpler problem is that of \emph{hybrid $k$-substring reconstruction}, in which the $k$-deck is replaced by the set of all \emph{substrings} of $\bfx$ of length $k$. This previously unexplored problem is relevant in the context of DNA sequence reconstruction from a combination of short (i.e., Illumina~\cite{illumina}) and long (i.e., Oxford Nanopore~\cite{oxford}) reads, and will be discussed elsewhere.    
 
The paper is organized as follows. In Section~\ref{sec:intro}, we introduce the problem and derive upper and non-asymptotic lower bounds on the hybrid $k$-deck size for the case than one long sequence is observed. In this setting, we show that under some constraints for $t$, we have $\log t + 2 < k \leq  \min  \{ t+1, O ( \sqrt{n \cdot (1+\log t)} )  \}$. 
For $t \leq 4$, we show that the upper bound is tight. We also consider the case of large $t$, in which case significantly smaller $k$-decks are needed for reconstruction. In Section~\ref{sec:multiK}, we consider the scenario when $M$ subsequences of $\bfx$ of length $n-t$ are available, along with the sequence's $k$-deck and describe a simple trace aggregation procedure that maps the problem to that of one asymmetric trace-aided reconstruction.

\section{Problem Statement and Single Trace Analysis} \label{sec:intro}

We introduce the hybrid $(t,k,M)$ $k$-deck problem, where one is asked to find the minimum value of $k$, denoted by $f(n,t,M)$, 
such that any binary sequence $\bfx$ may be reconstructed given $M$ subsequences $\cU = \{\underline{\bfx}_1, \ldots, \underline{\bfx}_M\}$ of $\bfx$ of length $n-t$ obtained by deleting zeros only, and the $k$-deck of $\bfx$ (note that the subsequences in the $k$-deck are obtained via deletions of both zeroes and ones). Clearly, we require that $k<n-t$, and mostly focus constant values of $t$ where $t=o(n)$. Nevertheless, we provide some results for the case $t=O(n)$ as well.
Furthermore, we start our analysis with the case $M=1$ and refer to the problem as the $(t,k)$ multi-deck problem. In this case, the goal is to find the minimum value of $k$, denoted by $f(n,t)$, 
such that reconstruction is possible given a \emph{single} length $n-t$ subsequence $\underline{\bfx}$ of $\bfx$ obtained by deleting zeros only, and the $k$-deck of $\bfx$. 

\begin{example}\label{ex:1} Suppose that $\bfx = (1,1,1,\textcolor{red}{0})$ and that $\underline{\bfx} = (1,1,1)$ is the observed subsequence $\underline{\bfx}$ of $\bfx$ of length $n-1=3$. In this case, we may reconstruct $\bfx$ given $\underline{\bfx}$ and the $2$-deck of $\bfx$, denoted by $\cX$,  \vspace{-0.5ex}
$$\Big \{ (1,1), (1,1), (1,\textcolor{red}{0}), (1,1), (1, \textcolor{red}{0})\Big \}. \vspace{-0.5ex}$$
(Observe that given the $k$-deck, one can uniquely reconstruct the $\ell$-decks for any $\ell <k$.)
Note that reconstructing $\bfx$ is straightforward since we know that only symbols of value $0$ may have be deleted: Since $(1, \textcolor{red}{0})$ appears three times in $\cX$, 
it follows that to obtain $\bfx$ from $\underline{\bfx}$ we need to insert $0$ in the last position of $\underline{\bfx}$. The $1$-deck does not suffice for reconstruction.
\end{example}

The following claim formalizes the above observation and establishes a connection between Varshamov-Tenengoltz (VT) codes~\cite{Sloane,Sala} and the $f(n,1)$ hybrid $k$-deck problem.

\begin{claim} For any positive integer $n \geq 2$, $f(n,1) \leq 2$. \end{claim}
\begin{IEEEproof} Following the approach of~\cite{Scott}, let $n_i$ denote the number of subsequences of $\bfx=(x_1,\ldots,x_n)$ of length $i$ that end with a one. Then, \vspace{-0.5ex}
$$n_i = \sum_{j=1}^n \nchoosek{j-1}{i-1} \cdot x_j. \vspace{-0.5ex}$$ 
In particular, we are interested in $i \in \{1,2\}$, in which case $n_1 = \sum_{j=1}^n x_j$ and $n_2 = \sum_{j=1}^n (j-1) \cdot x_j$. Let \vspace{-0.5ex}
$$S(\bfx) =n_1 + n_2 = \sum_{j=1} j \cdot x_j, \vspace{-0.5ex}$$ 
and set $a = S(\bfx) \bmod (n+1)$. Thus, $\bfx \in \cC(n,a)$ where $\cC(n,a) = \{ \bfx : \sum_{i=1}^n i \cdot x_i \equiv a \bmod (n+1) \}.$ It is known from \cite{Sloane} that $\cC(n,a)$ is a code capable of correcting a single deletion so that there exists a decoder for $\cC(n,a)$ that can uniquely determine $\bfx$ given $\underline{\bfx}$ and $a$. This proves the claim.
\end{IEEEproof}
\vspace{-0.1in}
\begin{corollary}\label{lem:1k} For a positive integer $n \geq 2$, $f(n,1) = 2$. \end{corollary}

\begin{theorem}\label{th:ub1} For positive integers $n\geq 2$ and $t<n$, one has $f(n,t) \leq t+1$.  \end{theorem} \vspace{-0.5ex}
\begin{IEEEproof}\vspace{-0.5ex} Let $\cX$ denote the $(t+1)$-deck of $\bfx$ and let $\underline{\cX}$ denote the $(t+1)$-deck of $\underline{\bfx}$. For $j \in [t]$, let  $n_{\bfx, 1^{j}0}$ denote the number of subsequences in $\cX$ that start with $j$ ones and end with a zero, and similarly, let $n_{\underline{\bfx}, 1^{j}0}$ denote the number of subsequences in $\underline{\cX}$ that start with $j$ ones and end with a zero. Suppose that $I(\bfx,\underline{\bfx})=\{{k_1, k_2, \ldots, k_t\}},$ where $k_1 < k_2 < \cdots < k_t$ correspond to the positions of the zeros deleted in $\bfx$ that lead to $\underline{\bfx}$ (For simplicity, we omit the arguments of $I(\bfx,\underline{\bfx})$ whenever the meaning is clear from the context).  As an example, if $I = \{ 1,3 \}$ and $\bfx = ( \textcolor{red}{0},0,\textcolor{red}{0}, 1,0)$, then $\underline{\bfx} = (0,1,0)$. For an integer $m \leq n$, let $1_{{\bfx}}(m)$ denote the number of ones that appear in ${\bfx}$ before position $m$. For example, if ${\bfx} = (0,0,0,1,0)$, then $1_{{\bfx}}(2) = 0$ and $1_{{\bfx}}(5) = 1$. 

Next, note that the difference $n_{\bfx, 1^j0} - n_{\underline{\bfx}, 1^j 0}$ equals \vspace{-0.5ex}
$$\nchoosek{1_{{\bfx}}(k_1)}{j} + \nchoosek{1_{{\bfx}}(k_2)}{j} + \cdots + \nchoosek{1_{{\bfx}}(k_t)}{j}, \vspace{-0.5ex}$$
as deleting a zero at position $k_i$ reduces the count of the $n_{\underline{\bfx}, 1^j0}$ sequences compared to $n_{\bfx, 1^j0}$ by $\nchoosek{1_{{\bfx}}(k_i)}{j}.$

Let $R=\Big \{ 1_{{\bfx}}(k_1), \ldots, 1_{{\bfx}}(k_t) \Big \}$ and let $F(x)$ be a polynomial with its set of roots equal to $R$. 
It is straightforward to see that given $n_{\bfx, 1^j0} - n_{\underline{\bfx}, 1^j 0},$ for $1 \leq j \leq t$, we may uniquely recover the the $j$-th power sum symmetric polynomials over $R$ recursively. Recall that the $j$-th power sum symmetric polynomial over the variables $a_1,a_2,\ldots,a_m$ is defined as \vspace{-0.5ex}
$$ p_j(a_1,\ldots,a_m)=\sum_{i=1}^m a_i^j. \vspace{-0.5ex}$$
Using Newton's identities~\cite{R06} one may evaluate the elementary symmetric polynomials $e_i, \, i=1,\ldots,t,$ over $R$ based on the power sum symmetric polynomials over $R$. The elementary symmetric polynomials are defined as
$$ e_0(R)=1, e_1(R)=1_{{\bfx}}(k_1)+\ldots+1_{{\bfx}}(k_t),\ldots$$
$$ e_{t-1}(R)=\sum_{i_1<i_2< \ldots <i_{t-1}} 1_{{\bfx}}(k_{i_1})\cdots 1_{{\bfx}}(k_{i_{t-1}}),$$ 
$$e_t(R)=1_{{\bfx}}(k_1)\cdots1_{{\bfx}}(k_t).$$
Thus, we can recover the polynomial $F(x)$ and the elements of $R$. This allows us to determine $\bfx$ from $R$ and $\underline{\bfx}$.
\end{IEEEproof}

We now turn our attention to lower bounds. We use the following notation: For a vector $\bfv \in \{0,1\}^n$, we let $\cD_t(\bfv) \subseteq \mathbb\{0,1\}^{n-t}$ denote the set of all sequences that may be obtained by deleting $t$ zeros from $\bfv$. Also, for a $\bfv' \in \cD_t(\bfv)$, we say that $\bfv'$ is an asymmetric subsequence (or subsequence for short) of $\bfv$ and that $\bfv$ is an asymmetric supersequence (or supersequence for short) of $\bfv'$. 



\begin{lemma}\label{lem:fint} For all positive integers $n\geq 2$ and $t<n$, one has $f(2n,2t) \geq f(n,t ) + 1$. \end{lemma}
\begin{IEEEproof} Assume that $f(n,t) = k+1$. Then, there exist two distinct binary vectors $\bfx, \bfy \in \{0,1\}^n$ with the same $k$-deck and such that $\underline{\bfx} \in \cD_t(\bfx)$ and $\underline{\bfx} \in \cD_t(\bfy)$. From \cite{MMSSS91}, we have that the $(k+1)$-deck of $\bfx \bfy$ is equal to the $(k+1)$-deck of $\bfy \bfx$. Clearly, $\underline{\bfx} \underline{\bfx} \in \cD_{2t}(\bfx \bfy)$ and $\underline{\bfx} \underline{\bfx} \in \cD_{2t}(\bfy \bfx)$. Thus, we have two sequences $\bfx \bfy$ and $\bfy \bfx,$ each of length $2n,$ sharing the same $(k+1)$-deck and containing the subsequence $\underline{\bfx} \underline{\bfx}$ of length $2n-2t$ Therefore, $f(2n, 2t) \geq k+2 = f(n,t) + 1,$ as desired.
\end{IEEEproof}
%
\vspace{-0.1in}
\begin{theorem}\label{th:ft} For $t \leq \frac{n}{2}$, $f(n,t) \geq \log t + 2$. \end{theorem}
\begin{IEEEproof} Let $\bfx = 01$ and $\bfy = 10$. Then, $f(2,1) \geq 2$ and from repeated application of Lemma~\ref{lem:fint}, we have $f(2^s, 2^{s-1}) \geq s+1.$ This establishes the claim. (For a related use of the infinite Morse-Thue sequence and its complement, the interested reader is referred to~\cite{DS02}).

\end{IEEEproof}
Using Theorem~\ref{th:ft}, we show next that the upper bound of Theorem~\ref{th:ub1} is tight for $t \leq 4$.

\begin{corollary} For $t \leq 4$, $f(n,t) = t+1,$ provided that $n \geq 2t$. \end{corollary}
\begin{IEEEproof} The claim for $t=1$ follows from Lemma~\ref{lem:1k}. The previous theorem established the result for $t=2$. The claim for $t=3$ follows by observing that $\bfx=(0,1,1,0,1,0,0,1)$ and $\bfy=(1,0,0,1,0,1,1,0)$ share a common supersequence of length $11$ and have the same $3$-deck. For $t=4$, the bound follows from the existence of two sequences - $(1,1,0,0,1,1,1,0,1,1,0,0,1)$ and $(1,0,1,1,1,0,1,0,0,1,1,1,0)$ - which share a common length $9$ subsequence and have the same $4$-deck.
\end{IEEEproof}
Let $N=1+wt(\bfx)$, where $wt(\bfx)$ denotes the weight of the vector $\bfx$. The next lemma provides an improvement of the result of Theorem~\ref{th:ub1} for the case that $t = N^{\epsilon}$ and $1/2 <\epsilon<1$. Similar to~\cite{P14}, we make use of the following result from \cite{BEK99}.

\begin{lemma}\label{lem:math} \textit{(c.f., \cite{BEK99})} There is an absolute constant $c > 0$ such that every polynomial $p$ of the form: 
$$ p(x) = \sum_{j=0}^n a_j \cdot x^j, |a_j| \leq 1, a_j \in \mathbb{C},$$
has at most $c \sqrt{n(1-\log |a_0|)}$ zeros at one.
\end{lemma}

\begin{theorem}\label{th:UBG} If $t = N^{\epsilon}$, where $1/2 <\epsilon<1$, than any sequence $\bfx \in \{0,1\}^n$ may be reconstructed given an asymmetric $n-t$ trace and a $k$-deck of $\bfx$ with
$$ k \leq  c \sqrt{N \cdot(1 + \epsilon \log N) },$$
where $c$ is a constant.
\end{theorem}
\begin{IEEEproof} The result follows by counting the number of subsequences from the $k$-deck that start with $j$ ones, for $j+1 \in [k]$, and end with a zero, denoted by $1^j 0$. For $b \in \{0,1\}$, let $\bar{b}=1-b$ denote its complement and assume that $\bfx = (x_1, \ldots, x_n)$. Furthermore, suppose that $\bfx$ has $wt(\bfx)$ ones and recall that $N = wt(\bfx)+1$. Let $\bfX=(X_1, \ldots, X_{N}) \in \{0,1,\ldots,n\}^{N}$ be a vector with elements defined as follows: For $i \in [N]$, $X_i$ equals the number of zeros between the $(i-1)$-th and $i$-th one in $\bfx$ (We tacitly assume that a one is pre-pended and a one is appended to the sequence first). For example, if $\bfx = (0,1,1,0)$, then $\bfX=(1,0,1)$. 

Note that similarly to our previous approach, we may write 
$$n_{\bfx, 1^j0} = \sum_{\ell=1}^n \nchoosek{1_{{\bfx}}(\ell)}{j} \cdot \bar{x}_\ell =  \sum_{\ell=1}^{N} \nchoosek{\ell-1}{j} \cdot X_\ell.$$
By linearly combining the counts $n_{\bfx, 1^j0}$ for different values of $j$ we can determine
$$s_j(\bfx)  = \sum_{\ell=1}^N \ell^{j} \cdot X_\ell. $$

Suppose next that $\bfu \in \{0,1\}^n$, $\bfu \neq \bfx$, and let $\bfx$ and $\bfu$ have the same $k$-deck. In addition, assume that there exists a sequence $\bfy \in \{0,1\}^{n-t}$ such that $\bfy \in \cD_t(\bfx)$ and $\bfy \in \cD_t(\bfu)$. Define $\bfU$ in a manner analogous to $\bfX$. Then
\begin{align}\label{eq:equalUPNE}
s_j(\bfx) = \sum_{\ell=1}^N \ell^j \cdot X_\ell =   \sum_{\ell=1}^N \ell^j \cdot U_\ell = s_j(\bfu), 
\end{align}
for $1 \leq j \leq k-1$. Let 
\begin{align*}
p_\bfx(z) = \sum_{\ell=0}^N X_\ell \cdot z^\ell,\ \ \  p_\bfu(z) = \sum_{\ell=0}^N U_\ell \cdot z^\ell.
\end{align*}
Furthermore, let $\left( \frac{\partial^j}{\partial z^j} p_{\bfx}(z) \right)_{z=1}$ be the $j$-th partial derivative of $p_\bfx(z)$ evaluated at $z=1$. Note that if (\ref{eq:equalUPNE}) holds, then 
$$\left( \frac{\partial^j}{\partial z^j} p_{\bfx}(z) \right)_{z=1} = \left( \frac{\partial^j}{\partial z^j} p_{\bfu}(z) \right)_{z=1}$$ 
holds as well. Letting $P(z) = p_{\bfx}(z) - p_{\bfu}(z)$, we have
$$ (1-z)^{k_m}  | P(z). $$
Assume that the degree of the polynomial $P(z)$ is $d$ and observe that for any $1 \leq \ell \leq N$, $|X_\ell - U_\ell| \leq t$, since by assumption, there exists a $\bfy$ such that $\bfy \in \cD_t(\bfx)$ and $\bfy \in \cD_t(\bfu)$. Define $f(z) = \frac{1}{z^d t} \cdot P(z)$; $f(z)$ satisfies the conditions of Lemma~\ref{lem:math}, so that is has at most $c \sqrt{N \cdot (1-\log |\frac{1}{t}|)}$ zeros at one, which implies
$$k_m \leq  c \sqrt{N \cdot (1+\log t)}.$$ 
Substituting $t=N^\epsilon$ proves the claim.
\end{IEEEproof}
The previous result improves upon Theorem~\ref{th:ub1} for the case when $\epsilon > \frac{1}{2}$. For large values of $t$, an alternative approach is to discard the vector $\underline{\bfx}$ and reconstruct $\bfx$ using only the $k$-deck for $\bfx$ according to~\cite{KR73}, \cite{Scott}. For the case when $n >> N$, Theorem~\ref{th:UBG} improves upon the best known result in the literature~\cite{KR73}, which asserts that $f(n,n) \leq (1+o(1)) \frac{16}{7} \sqrt{n}$. 
%
The following corollary summarizes Theorem~\ref{th:ub1} and Theorem~\ref{th:UBG}.

\begin{corollary}\label{cor:UB}  For $\bfx \in \{{0,1\}}^n$ such that $wt(\bfx) = N-1$ and $t=N^\epsilon,$ where $1/2 < \epsilon <1$,
$$f(n,t) \leq \min \Big \{ N^{\epsilon}+1, O \Big( \sqrt{N \cdot (1+ \epsilon \log N)} \Big) \Big \}. $$
\end{corollary}

\section{The Multitrace Reconstruction Problem}\label{sec:multiK}

We focus next on the scenario where one is given $M$ trace sequences $\cU = \{ \underline{\bfx}^{(1)}, \underline{\bfx}^{(2)},$ $\ldots, \underline{\bfx}^{(M)} \}$ of length $n-t$, each of which is obtained by deleting $t$ zeros from $\bfx$. The question of interest is to determine the minimum value of $k$, denoted by $f(n,t,M)$, such that it is possible to reconstruct $\bfx$ given the set $\cU$ along with the $k$-deck of $\bfx$.

For a set $S \subseteq \{0,1\}^m$ and a sequence $\bfv \in \{0,1\}^k$, let $\bfv \circ S$ denote the set obtained by pre-pending to every element in $S$ the vector $\bfv$. For instance if $S=\{ (0,1), (1,1) \}$ and $\bfv = (0,0)$, then $\bfv \circ S = \{ (0,0,0,1), (0,0,1,1) \}$. For a vector $\bfv \in \{0,1\}^n$, let $\cI_t(\bfv)$ denote the set of vectors that may be obtained by inserting $t$ zeros into $\bfv$. For instance, if $\bfv = (0,1)$, then $\cI_1(\bfv) = \{ (\textcolor{red}{0},0,1), (0,1,\textcolor{red}{0}) \}$.

\begin{lemma} For positive integers, $r \geq 2,t>1,1 \leq M < n-1,$  
$$f(2r,r-M+1,M) \geq f(2(r-M), r-M),$$
and
$$ f(2r+1,r-M+1,M+1) \geq f(2(r-M), r-M). $$
\end{lemma}
\begin{IEEEproof} Let $\bfa = (0,1,0,1,\ldots,0,1) \in \{0,1\}^{2M}$ and suppose that we have two sequences $\bfx = (\bfx', \bfa) \in \{0,1\}^{2T+2M}$ and $\bfy = (\bfy', \bfa) \in \{0,1\}^{2T + 2M}$ such that $\bfx', \bfy' \in \{{0,1\}}^{2T}$ have the same $k$-deck and such that there exists a $\bfz \in \cD_{T}(\bfx') \cap \cD_{T}(\bfy')$ (i.e, $\bfx'$ and $\bfy'$ share a trace of length $T$). Clearly, under this setup, $\bfx, \bfy$ have the same $k$-deck. 

First, note that $\bfz \circ \cD_{1}(\bfa) \subseteq \cD_{T+1}(\bfx)$.  
Since $\bfz \in \cD_T(\bfy')$, we also have $ \bfz \circ \cD_{1}(\bfa) \subseteq \cD_{T+1}(\bfy)$. Furthermore, since $|D_1(\bfa)| \geq M$, one also has $|\bfz \circ \cD_{1}(\bfa)| \geq M$. Let $\cU_{\bfz} \subseteq \bfz \circ \cD_{1}(\bfa)$, say $\cU_{\bfz}= \{ \underline{\bfx}^{(1)}, \underline{\bfx}^{(2)},$ $\ldots, \underline{\bfx}^{(M)} \}$. Then, $\bfx, \bfy$ are such that for all $i \in [M]$, $\bfx,\bfy \in \cI_{T+1}(\underline{\bfx}^{(i)})$. Thus, $f(2T + 2M, T+1, M) \geq f(2T,T)$. The statement in the lemma follows now by setting $r=M+T$.

For the case that $\bfx$ and $\bfy$ have odd length, we let the alternating sequence $\bfa$ have length $2M+1$, and $|\cU_{\bfz}| = M + 1$. In this case, we get $f(2T + 2M + 1, T+1, M+1) \geq f(2T, T)$. Substituting $r=M+T$ gives the second expression.
\end{IEEEproof}

\begin{example} Suppose that $M=3$ and that $T=2$. Let $\bfx' = (0,1,1,0)$, $\bfy' = (1,0,0,1)$, $\bfa = (0,1,0,1,0,1)$ and observe that $\bfz = (1,1)$ is a common subsequence of both $\bfx'$ and $\bfy'$. Then, we may choose $\cU=\{ (1,1,1,0,1,0,1), (1,1,0,1,1,0,1),\\ (1,1,0,1,0,1,1) \} = \{ \underline{\bfx}^{(1)}, \underline{\bfx}^{(2)}, \underline{\bfx}^{(3)} \}$,  such that $\cU \subseteq \cD_T(\bfx)$ and $\cU \subseteq \cD_T(\bfy),$ where $\bfx = (0,1,1,0,0,1,0,1,0,1)$ and $\bfy =  (1,0,0,1, 0,1,0,1,0,1)$. Thus, we have $3$ sequences each of length $10-3=7$ where each sequence is a subsequence of both $\bfx$ and $\bfy$. Since $\bfx$ and $\bfy$ have the same $2$-deck, it follows from that $f(2\cdot 2 + 2 \cdot 3, 3, 3)=f(8,3,3) \geq f(4,2) = 3$. \end{example}

We now turn our attention to an upper bound. Let $N = wt(\bfx) + 1$ and $\bfX = (X_1, \ldots, X_N)$ be as defined in the previous lemmas. In addition, reserve $\underline{\bfX}^{(m)}$, $1 \leq m \leq M$ for the sequence $\bfX$ of $\bfx^{(m)}$ obtained by counting the occurrences of zeros between ones as described in the proof of Theorem~\ref{th:UBG}.

\begin{lemma} For positive integers $n \geq 2,t<n$, and $M \geq 1$, $f(n,t,M) \leq f(n,t-1) \leq t$. \end{lemma}
\begin{IEEEproof} Suppose that $f(n,t,M) \leq f(n,t)$, and let $M =2$ and $\cU = \{ \bfx^{(1)}, \bfx^{(2)} \}$. Observe that $f(n,t,M)$ is non-increasing in $M$, hence it suffices to analyze the case $M=2$ only. Furthermore, $d_H(\bfX^{(1)}, \bfX^{(2)}) \geq 1$ since otherwise $|\cU|=1$. Since $d_H(\bfX^{(1)}, \bfX^{(2)}) \geq 1$, we can identify and correct at least one deletion since we can find at least one run of zeros in $\bfx^{(1)}$ that underwent a deletion. Let $\underline{\bfx} \in \{0,1\}^{n-t+1}$ be the vector which results from correcting this deletion in $\bfx^{(1)}$. Then, the minimum $k$-deck required to reconstruct $\bfx$ given $\cU$ and $\underline{\bfx}$ is at most $f(n,t-1)$ which proves the statement in the lemma. 
\end{IEEEproof}
\begin{corollary} For $t \leq 5$, and $M \leq n-2t$, $$f(n,t,M) = t.$$ \end{corollary}
\begin{example} Suppose that $\bfx = (1,0,1,1,0,0,1,0)$ so that $\bfX = (0,1,0,2,1)$. Assume that we observe the following subsequences of length $n-t=n-2=6$ of $\bfx$, $\cU = \{ (1,1,1,0,1,0), (1,0,1,1,0,1) \}$. Hence, $\bfX^{(1)} = (0,\textcolor{red}{0},0,\textcolor{red}{1},1)$ and $\bfX^{(2)} = (0,1,0,\textcolor{red}{1},\textcolor{red}{0})$. Let $\bar{\bfX}=(\bar{X}_1, \bar{X}_2, \bar{X}_3, \bar{X}_4)$ be given according to $\bar{X}_i = \max \Big \{ X^{(1)}_i, X^{(2)}_i \Big \}$. Then, $\bar{\bfX} = (0,1,0,1,1)$ and $\bar{\bfx} =(1,0,1,1,0,1,0)$. Note that $d_H(\bfX, \bar{\bfX}) = 1$ and that $\bar{\bfx}$ is the result of deleting a zero from $\bfx$. Let $n_{1 0, \bfx}$ denote the number of occurrences of the subsequence $10$ in $\bfx$ and similarly, let $n_{10, \bar{\bfx}}$ denote the number of occurrences of the subsequence $10$ in $\bar{\bfx}$. Since $n_{10,\bfx} - n_{10, \bar{\bfx}} = 11 - 8=3$, we need to add one to the value at the third position of $\bar{\bfX}$ to obtain $\bfX$. From $\bfX$, we can then recover $\bfx$. 
\end{example}
Next, we consider the case when $M$ is sufficiently large to guarantee a signifiant reduction in the value of the deck length $k$. In our proofs, we make use of the following claims.
\begin{claim}\label{cl:clINT} Let $\bfx \in \{0,1\}^n$, $\bfy \in \{0,1\}^n$ be such that there exists a $\bfw \in \{0,1\}^{n+t},$ such that $\bfw \in \cI_t(\bfx) \cap \cI_t(\bfy)$. Let $t_0 \leq t$ be the smallest possible integer for which $\cI_{t_0}(\bfx) \cap \cI_{t_0}(\bfy) \neq \emptyset$ and suppose that $\bfz \in \cI_{t_0}(\bfx) \cap \cI_{t_0}(\bfy)$. Then, $\bfw \in \cI_{t-t_0}(\bfz)$. \end{claim}
\begin{IEEEproof} The result follows by noting that for any two strings $\bfv, \bfw$ such that $\bfw \in \cI_t(\bfv)$, we have $W_i \geq V_i$ for $i \in [N]$. Here, $\bfV=(V_1, \ldots, V_N)$ and $\bfW=(W_1, \ldots, W_N)$ denote the $\bfX$-analogues of $\bfv$ and $\bfw$. 
\end{IEEEproof}

\begin{example} Suppose that $\bfx = (0,1,1,0,1)$ and $\bfy = (0,0,1,1,1)$ so that $\bfX = (1,0,1,0)$ and $\bfY = (2,0,0,0)$. Then $\bfZ$ may be formed by taking the maximum element of $\bfX = (X_1, \ldots, X_4)$ and $\bfY = (Y_1, \ldots, Y_4)$,  $\bfZ = (2,0,1,0) = (Z_1, \ldots, Z_4)$. This gives $\bfz = (0,0,1,1,0,1)$. Observe that if $\bfw$ is any asymmetric supersequence of $\bfX$ and $\bfY$, then for $i \in [4]$, we require $W_i \geq X_i$ and similarly $W_i \geq Y_i$ which implies that $W_i \geq Z_i,$ since $Z_i = \max\{ X_i, Y_i \}$.
\end{example}

\begin{claim}\label{cl:maxball} Suppose that $n \geq 2$. Then, for $t < \lfloor \frac{n}{6} \rfloor$, one has
$$\max_{\bfz \in \{0,1\}^n} |\cD_t(\bfz)| \leq \nchoosek{\lceil \frac{n}{2} \rceil}{t}.$$ \end{claim}
\begin{IEEEproof} Let $\bfa = (0,1,0,1,0,1, \ldots, ) \in \{0,1\}^m$ be the alternating string of length $m,$ and suppose that $\bfv \in \{0,1\}^m,$ $\bfv \neq \bfa$, is an arbitrary binary string of length $m$ that contains at least one run of zeros of length $1$ (i.e., the substring $101$). 

We first show that $|\cD_t(\bfa)| \geq |\cD_t(\bfv)|$ when $t \leq \lceil \frac{m}{2} \rceil$. The proof proceeds by induction. We first establish the base case. For $t=1$ and for an arbitrary $m$, $|\cD_t(\bfa)| \geq |\cD_t(\bfv)|$. Furthermore, for any $t \leq \lceil \frac{m}{2} \rceil$, it is straightforward to see $|\cD_t(\bfa)| \geq | \cD_t(\bfv) |$ since $\bfv$ has at most $t=\lceil \frac{m}{2} \rceil$ runs of zeros. Next, for the inductive step, suppose that $m+t = s$ and assume that the claim holds for all $m+t < s$. 
Suppose the first occurrence of $101$ in $\bfv$ from the left starts at position $j$. We partition the set $\cD_t(\bfv)$ as follows:
\begin{itemize}
\item $\cD(\bfv)^{(0)}$: The set of all sequences in $\cD_t(\bfv)$ in which the zero between the positions $j$ and $(j+2)$ is not deleted. 
\item $\cD(\bfv)^{(1)}$: The set of all sequences in $\cD_t(\bfv)$ in which the zero between the positions $j$ and $(j+2)$ is deleted. 
\end{itemize}
We partition the set $\cD_t(\bfa)$ similarly:
\begin{itemize}
\item $\cD(\bfa)^{(0)}$: The set of sequences in $\cD_t(\bfa)$ that start with zero.
\item $\cD(\bfa)^{(1)}$: The set of sequences in $\cD_t(\bfa)$ that start with one.
\end{itemize}
Note that $|\cD(\bfa)^{(0)}| = |\cD_t(\bfa')|,$ where $\bfa' = (0,1,0,\ldots) \in \{0,1\}^{m-2},$ and that $|\cD(\bfa)^{(1)}| =  |\cD_{t-1}(\bfa')|$. Also, $|\cD(\bfv)^{(0)}| = |\cD_{t}(\bfv')|,$ where $\bfv'$ is the length $m-2$ sequence obtained by deleting the string $10$ starting at index $j$ from $\bfv$. In addition, $|\cD(\bfv)^{(1)}| = |\cD_{t-1}(\bfv')|$. Since $m-2+t<s$, can apply the inductive hypothesis to determine $|\cD_t(\bfa')| \geq |\cD_t(\bfv')|$ and $|\cD_{t-1}(\bfa')| \geq |\cD_{t-1}(\bfv')|$, which implies $|\cD_t(\bfa)| \geq |\cD_t(\bfv)|$ when $m+t = s$.

Consider next the case when $\bfv$ is any length-$n$ vector that has no runs of zeros of length one, and let $t < \lfloor n/6 \rfloor$. In this case, $|\cD_t(\bfv)| \leq \left( n/3 \right)^t$ since $\bfv$ has at most $n/3$ runs of zeros, and $|\cD_t(\bfa)| \geq \nchoosek{ \lfloor n/2 \rfloor }{t}$. Since $\nchoosek{ \lfloor n/2 \rfloor }{t} \geq \left( n/3 \right)^t$ when $t < \lfloor n/6 \rfloor$, the result follows.
\end{IEEEproof}

Using the previous claims, we can establish upper and lower bounds on $f(n,t,M)$.

\begin{lemma} For integers $n\geq 2,t<n,M \geq 1$, let $m_0 = \Big \lfloor \frac{\log M}{\log n} + (n-t) \Big \rfloor$. Then, for $t < \lfloor \frac{m_0}{6} \rfloor$,
$$ f(n,t,M) \leq f( n, n-m_0). $$
\end{lemma}
\begin{IEEEproof} Under the assumptions of Claim~\ref{cl:clINT} applied to $M$ sequences, we seek the smallest possible length sequence $\bfz \in \{0,1\}^m$, $m \geq n-t$, such that $\bfz \in \cI_{m-n+t}(\underline{\bfx}^{(1)}) \cap \cI_{m-n+t}(\underline{\bfx}^{(2)}) \cap \cdots \cap \cI_{m-n+t}(\underline{\bfx}^{(M)})$. According to Claim~\ref{cl:maxball}, for $t < \lfloor \frac{m}{6} \rfloor$ we have
$$ | \cD_{t-(n-m)}(\bfz) | \leq \nchoosek{\lceil \frac{m}{2} \rceil}{t-(n-m)}.$$
Since $\nchoosek{\lceil \frac{m}{2} \rceil}{t-n+m} \leq \left( \lceil \frac{m}{2} \rceil \right)^{t-n+m} $, if 
$$ m = m_0= \Big \lfloor  \frac{\log M}{\log n} + (n-t) \Big \rfloor, $$
then 
$$ M > |\cD_{t-(n-m)}(\bfz)|. $$
Hence, $\bfz$ has length at least $m$ and $\bfz \in   \cI_{m-n+t}(\underline{\bfx}^{(1)}) \cap \cI_{m-n+t}(\underline{\bfx}^{(2)}) \cap \cdots \cap \cI_{m-n+t}(\underline{\bfx}^{(M)})$. We can determine the sequence $\bfx$ given the length $n-m$ subsequence $\bfz \in \cI_{n-m}(\bfx)$ and its $f(n,n-m)$-deck.
\end{IEEEproof}

\begin{lemma} For integers $n\geq 2,t,M \geq 1$, let $m = \Big \lceil  \frac{\log M}{-\log \left({2(1-\frac{n-t}{n-t+1})}\right)} + (n-t) \Big \rceil$. Then, for $t < \lfloor \frac{m}{6} \rfloor$,
$$ f(n,t,M) \geq f( n, n-m). $$
\end{lemma} 
\begin{IEEEproof} Under the assumptions of Claim~\ref{cl:clINT} applied to $M$ sequences, we need to determine the minimum length sequence $\bfz \in \{0,1\}^m$, $m\geq n-t$, such that $\bfz \in \cI_{m-n+t}(\underline{\bfx}^{(1)}) \cap \cI_{m-n+t}(\underline{\bfx}^{(2)}) \cap \cdots \cap \cI_{m-n+t}(\underline{\bfx}^{(M)})$. Wlog, assume that $\bfz$ is the alternating string. Then, $ | \cD_{t-(n-m)}(\bfz) | \geq \left( \frac{m/2}{t-n+m} \right)^{t-n+m} = \left( \frac{1}{2(1-\frac{n-t}{m})} \right)^{t-n+m}$. Since $m \geq n-t+1$, if $m = \Big \lceil  \frac{\log M}{-\log \left({2(1-\frac{n-t}{n-t+1})}\right)} + (n-t) \Big \rceil$, then  $M<|\cD_{t-(n-m)}(\bfz)|$. Hence, $f(n,t,M) \geq f(n,n-m)$.
\end{IEEEproof}
\begin{theorem}\label{th:UBLB} For integers $n\geq 2,t<n,M\geq 1$,  
$$ f(n,t,M) = f( n, n-m)$$
where $\Big \lfloor \frac{\log M}{\log n} + (n-t) \Big \rfloor \leq m \leq\Big \lceil  \frac{\log M}{-\log \left({2(1-\frac{n-t}{n-t+1})}\right)} + (n-t) \Big \rceil$, and $t < \lfloor \frac{m}{6} \rfloor$.
 \end{theorem}
Invoking the results of the previous section, we arrive at the following corollary.
\begin{corollary} Suppose that $t < \lfloor \frac{m}{6} \rfloor$ and $M = \nchoosek{\frac{m}{2}}{t-n+m} + 1$ where $m$ is an even integer. If $n-m \leq 4$, then 
$$ f(n,t,M) = n-m+1.$$
\end{corollary}
\small
\textbf{Acknowledgement.} This research was supported in part by the NSF grants CIF CCS 1526875 and 1618366, and the NSF STC Center for Science of Information at Purdue University.


\vspace{-0.1in}
\begin{thebibliography}{1}

\bibitem{acharya2015} J. Acharya, H. Das, O. Milenkovic, A. Orlitsky, and S. Pan, ``String reconstruction from substring compositions,'' \emph{SIAM Journal on Discrete Mathematics} 29, no. 3, 1340-1371, 2015.

\bibitem{batu2004} Batu, Tukan, Sampath Kannan, Sanjeev Khanna, and Andrew McGregor. "Reconstructing strings from random traces." In Proceedings of the fifteenth annual ACM-SIAM symposium on Discrete algorithms, pp. 910-918. Society for Industrial and Applied Mathematics, 2004.

\bibitem{bondy77} Bondy, John Adrian, and Robert L. Hemminger. "Graph reconstruction a survey." Journal of Graph Theory 1, no. 3 (1977): 227-268.

\bibitem{BEK99} P. Borwein, T. Erdelyi, G. Kos, ``Littlewood-type problems on [0,1],'' \textit{Proc. London Math. Soc.}, vol. 79, no. 1, pp. 22-46, 1999.

\bibitem{CK97} C. Choffrut and J. Karhumaki, ``Combinatorics of words,'' in \textit{Handbook of Formal Languages}, vol. I, Springer, Berlin, 1997, pp. 329-438.

\bibitem{DS02} M. Dudik and L.J. Schulman, ``Reconstruction from subsequences,'' \textit{Journal of Combinatorial Theory}, vol. 103, no. 2, pp. 337-348, 2003.

\bibitem{GY15} R. Gabrys and E. Yaakobi, ``Sequence reconstruction over the deletion channel,'' \textit{Proc. IEEE ISIT}, Barcelona, 2016.

\bibitem{kalashnik73} Kalashnik, L. O. ``The reconstruction of a word from fragments,'' \emph{Numerical Mathematics and Computer Technology}, Akad. Nauk. Ukrain. SSR Inst. Mat., Preprint IV (1973): 56-57.

\bibitem{KR73} I. Krasikov and Y. Roditty, ``On a reconstruction problem for sequences,'' \textit{Journal of Combinatorial Theory}, vol. 77, no. 2, pp. 344-348, 1997.  

\bibitem{MMSSS91} B. Manvel, A. Meyerowitz, A. Schwenk, K. Smith, and P. Stockmeyer, ``Reconstruction of sequences,'' \textit{Discrete Math}, vol. 94, no. 3, pp. 209-219, 1991.

\bibitem{illumina} Mardis, Elaine R. "The impact of next-generation sequencing technology on genetics." Trends in genetics 24, no. 3 (2008): 133-141.

\bibitem{oxford} Oxford Nanopore Technologies, ``DNA: nanopore sequencing,'' available at \textit{https://nanoporetech.com/applications/dna-nanopore-sequencing}, 2017.

\bibitem{P14} {A. Pacchiano}, \textit{Trace Reconstruction Problem} (Master's Thesis), retrieved from \textit{http://dspace.mit.edu/bitstream/handle/1721.1/91856/894352537-MIT.pdf?sequence=2}, 2014. 

\bibitem{R06} {Roth, R.,} \textit{Introduction to coding theory}, Cambridge University Press, 2006.

\bibitem{Sala} F. Sala, R. Gabrys, C. Schoeny, and L. Dolecek, ``Exact Reconstruction from Insertions in Synchronization Codes,'' 
to appear in \emph{IEEE Transactions on Information Theory}, 2017.

\bibitem{Scott} A.D. Scott, ``Reconstructing sequences,'' \textit{Discrete Mathematics}, vol. 175, no. 1-3, pp. 231-238, 1997.

\bibitem{Sloane} N.J.A. Sloane, ``On single-deletion-correcting codes,'' \textit{Codes and Designs - Ray-Chaudhuri Festschrift}, pp. 273-291, 2002. 

\bibitem{yazdi2016} Yazdi, SM Hossein Tabatabaei, Ryan Gabrys, and Olgica Milenkovic, ``Portable and Error-Free DNA-Based Data Storage,'' bioRxiv (2016): 079442.

\bibitem{zenkin1984} Zenkin AI, Leont'ev VK, ``On a non-classical recognition problem,'' \emph{USSR Computational Mathematics and Mathematical Physics}, 1984 Dec 31;24(3):189-93.


\end{thebibliography}
\end{document}